\documentclass[showpacs,aps,prd,preprint,nofootinbib,showkeys,unsortedaddress,raggedbottom]{revtex4-1}
\pdfoutput=1

\usepackage{bm}
\usepackage{amsmath}
\usepackage{graphicx}
\usepackage{subfigure}
\usepackage[usenames,dvipsnames]{color}
\definecolor{darkblue}{RGB}{0,0,196}
\usepackage[colorlinks=true,linkcolor=darkblue,citecolor=darkblue,urlcolor=darkblue]{hyperref}
\usepackage{slashed}

\usepackage{setspace}
\usepackage{footmisc}
\usepackage[makeroom]{cancel}

\newcommand{\intdP}{\int\!dP}

\def\be{\begin{equation}}
\def\ee{\end{equation}}
\def\ba{\begin{eqnarray}}
\def\ea{\end{eqnarray}}

\begin{document}

\title{Nonextensive hydrodynamics of boost-invariant plasmas}

\author{Mubarak Alqahtani} 
\affiliation{Department of Physics, College of Science, 
Imam Abdulrahman Bin Faisal University, Dammam 31441, Saudi Arabia }

\author{Nasser Demir} 
\affiliation{Department of Physics, Faculty of Science, Kuwait University, P.O. Box 5969, Safat 13060, Kuwait}

\author{Michael Strickland} 
\affiliation{Department of Physics, Kent State University, Kent, OH 44242 United States}

\begin{abstract}
 We use quasiparticle anisotropic hydrodynamics to study the non-conformal and non-extensive dynamics of a system undergoing boost-invariant Bjorken expansion.  To introduce nonextensivity, we use an underlying Tsallis distribution with a time-dependent nonextensivity parameter $q$.
 By taking moments of the quasiparticle Boltzmann equation in the relaxation-time approximation, we obtain dynamical equations which allow us to determine the time evolution of all microscopic parameters including $q$.  We compare numerical solutions for bulk observables obtained using the nonextensive evolution with results obtained using quasiparticle anisotropic hydrodynamics with a Boltzmann distribution function ($q \rightarrow 1$). We show that the evolution of the temperature, pressure ratio, and scaled energy density, are quite insensitive to which distribution function is assumed. However, we find significant differences in the early-time evolution of the bulk pressure which are observed for even small deviations from the Boltzmann distribution function. Finally, we discuss the existence of non-conformal hydrodynamic attractors for the longitudinal and transverse pressures, the bulk and shear viscous corrections, and the nonextensivity parameter $q$.
\end{abstract}

\date{\today}

\pacs{12.38.Mh, 24.10.Nz, 25.75.-q, 51.10.+y, 52.27.Ny}

\keywords{Quark-gluon plasma, Relativistic heavy-ion collisions, Anisotropic hydrodynamics, Tsallis statistics}

\maketitle

\section{Introduction}
\label{sec:intro}

Ultrarelativistic heavy-ion collision experiments allow physicists to study the behavior of nuclear matter at extremely high temperatures. In such experiments, high-energy collisions of heavy nuclei are used to heat a volume of matter up to temperatures that exceed the critical temperature ($T_c \sim  155$ MeV) necessary to create a quark-gluon plasma (QGP)~\cite{Florkowski:2010zz,Chaudhuri:2012yt,Averbeck:2015jja,1802.04801}. The bulk evolution of the QGP, as a strongly interacting state of matter, is well described by relativistic hydrodynamics~\cite{nucl-th/0305084,0902.3663,1301.5893,1605.08694,Jeon:2016uym,1707.02282,Romatschke:2017ejr}. In the last two decades, different frameworks have been developed to describe many heavy-ion collision observables~\cite{Huovinen:2001cy,Romatschke:2007mq,0911.2397,Niemi:2011ix,Ryu:2015vwa,1409.8164,Alqahtani:2017jwl,1711.08499,1901.01319} (see Ref.~\cite{2010.12377} for a recent review). One of the frameworks used is anisotropic hydrodynamics, which was introduced to take into account the fact that the QGP is a highly momentum anisotropic plasma at early times after the nuclear collision~\cite{1007.0130,1007.0889}. Motivation and the basics of anisotropic hydrodynamics are presented in detail in Ref.~\cite{1410.5786}.

In recent years, 3+1D quasiparticle anisotropic hydrodynamics has been developed where three momentum-space anisotropy parameters in the underlying distribution function and a single-finite thermal mass which is fit to lattice QCD data for the equation of state are included. The results from this model have been compared with a variety of heavy-ion observables at different collision energies: Au-Au collisions at 200 GeV and Pb-Pb collisions at 2.76 TeV and 5.02 TeV~\cite{1703.05808,1705.10191,1807.04337,2007.04209,2008.07657}. In all these cases, the 3+1d quasiparticle anisotropic hydrodynamics model was able to describe the data reasonably well for many heavy-ion observables such as the spectra, mean transverse momentum of identified hadrons, multiplicities, the elliptic flow, and the HBT radii. See Refs.~\cite{1712.03282,Alalawi:2021jwn} for recent reviews of 3+1D quasiparticle anisotropic hydrodynamics. 

One of the challenges of hydrodynamic models is to describe the spectra at intermediate transverse momentum, $p_T \sim 3$ GeV. As an example, in the 3+1D aHydroQP model, especially in peripheral collisions, one observes that the agreement between the aHydroQP model and experimental results is good only for $p_T  \lesssim 1.5$ GeV~\cite{2008.07657}. While the data exhibits a characteristic power-law tail, aHydroQP predictions exhibit an exponential behavior which falls faster than the experimental data does. This difference is primarily due to the fact that thermal distribution functions, which are asymptotically of exponential form, are assumed at freeze-out in aHydroQP. In contrast, when used at freeze-out, statistics based on a Tsallis distribution function can provide  exponential behavior at low $p_T$ and power-law behavior at high $p_T$~\cite{Tsallis:1987eu,Tsallis:1998ws,Tsallis:2003vv,Tsallis:2009zex}. 

Tsallis statistics has been used to fit the spectra more accurately over a wide range of $p_T$ in many high-energy experiments involving different systems and different collision energies,~\cite{nucl-th/9902070,nucl-ex/0607033,0710.1905,0810.2939,0812.1609,1007.0719,1005.3674,1012.5104,1110.5526,1203.4343,1905.12756,1908.04208,2003.03278,1912.01404,2010.14880,Sarwar:2021csp}. Rather than performing a direct fit to the data using a Tsallis distribution for the freeze-out hypersurface distribution, one would like to account for the nonextensivity in the underlying dynamical model, using Tsallis non-extensive statistics.  It would, therefore, be of interest to employ the Tsallis distribution function in the dynamical model itself to describe the characteristics of the medium in both the QGP and the freeze-out phases in order to study the effects on the predicted bulk observables.

In a prior paper~\cite{1509.02913}, quasiparticle anisotropic hydrodynamics was studied in 0+1D systems undergoing boost-invariant Bjorken expansion with the assumption of a Boltzmann distribution function. In this work, we derive the dynamical equations for the same systems using a Tsallis distribution function. After solving the dynamical equations, we compare the bulk observables predicted by this approach with the results presented before in Ref.~\cite{1509.02913}. We show that the evolution of temperature, pressure anisotropy, and energy density are not very sensitive to which distribution function is assumed in the aHydroQP framework. However, we find significant differences in the evolution of the bulk pressure which are observed for even slight deviations from the Boltzmann distribution function. We finally investigate the existence of dynamical attractors in both approaches. We present the evolution of the longitudinal and transverse pressures and the  bulk and shear viscous stresses as a function of time in units of the local relaxation time. We find that far-from-equilibrium non-conformal attractors exist for both extensive and non-extensive statistics.

The structure of the paper is as follows.  In Sec.~\ref{sec:distFuncs}, we introduce the Tsallis and Boltzmann distribution functions. In Sec.~\ref{sec:3+1daHydroQP}, we review the basics of the 0+1D quasiparticle anisotropic hydrodynamics model. In Sec.~\ref{sec:results}, results are presented for comparisons between the Boltzmann and Tsallis distribution functions using the 0+1D quasiparticle anisotropic hydrodynamics framework. Sec.~\ref{sec:conclusions} contains our conclusions and an outlook for the future. 


\section{Tsallis distribution function }
\label{sec:distFuncs}

The Tsallis non-extensive distribution function is given by~\cite{Tsallis:1987eu,Tsallis:1998ws,1608.08965}

\be
f_{T}(x,q)=[1+(q-1)x]^{\frac{-1}{q-1}}\,  \, \, \, \, {\rm with} \, x \geq 0 \, ,
\label{eq:TsallisDist1}
\ee
where $q$ is the Tsallis parameter which characterizes the nonextensivity of the system. By expanding Eq.~(\ref{eq:TsallisDist1}) around $q=1$, one obtains
\be
f_{T}(x,q)=e^{-x}+\frac{1}{2} \, (q-1) \, x^2 e^{-x}+... \,,
\label{eq:expandedTsallisDist}
\ee
where the leading term is the Boltzmann distribution function.  Since the $q \rightarrow 1$ limit yields the Boltzmann distribution, it is helpful to characterize the Tsallis distribution function in terms of deviations from $q=1$ (which would correspond to the Boltzmann limit).  For this purpose we introduce the notation $\delta q \equiv q-1$ which is a measure of the degree of non-extensivity, so that the $\delta q =0$ limit corresponds to the Boltzmann distribution. 

Hence the Tsallis distribution can be rewritten as 

\be
f_{T}(x,\delta q)=(1+\delta q \, x)^{-1/\delta q} \,.
\label{eq:TsallisDist2}
\ee
Henceforth all results obtained using the Tsallis distribution function shall be characterized in terms of the parameter $\delta q$, and all results must be interpreted in light of how far or near the system is from the thermal distribution. An interpretation of the Tsallis distribution and  a comparison to Boltzmann distribution with illustrations of the number density are provided in Refs.~\cite{0812.1471,hep-ph/9908459,0810.2939}. For more information concerning the application of Tsallis statistics to high-energy proton-proton, proton-nucleus, and nucleus-nucleus collisions, we refer the reader to the recent article of Kapusta~\cite{Kapusta:2021zfo} and references therein.

\section{Anisotropic hydrodynamics}
\label{sec:3+1daHydroQP}

In the canonical anisotropic hydrodynamics approach, the distribution function is assumed to be anisotropic in momentum space i.e, in the local rest frame one has~\cite{1007.0130,1007.0889,1509.02913}
\be
f(x,p) =  f_{\rm eq}\!\left(\frac{1}{\lambda}\sqrt{\sum_i \frac{p_i^2}{\alpha_i^2} + m^2}\right) ,
\label{eq:fform}
\ee
where $\lambda$ corresponds to a temperature-like parameter, $\alpha_i$ corresponds to the momentum anisotropy parameter in the $i$-th direction, and $f_{\rm eq}$ is an exponential Boltzmann distribution.  In the limit where $\alpha_i=1$ and $\lambda=T$, one recovers the isotropic thermal distribution.
In this work, the distribution function has the same argument as in Eq.~(\ref{eq:fform}), but with Tsallis form rather than Boltzmann form i.e.,
\be
f(x,p) =  f_{T}\!\left(\frac{1}{\lambda}\sqrt{\sum_i \frac{p_i^2}{\alpha_i^2} + m^2}\right) .
\label{eq:fformT}
\ee
Note that in the limit of $\delta q \rightarrow 0$, Eq.~(\ref{eq:fformT}) reproduces Eq.~(\ref{eq:fform}).
\subsection{Bulk Variables}
\label{subsec:bulkvars}
The bulk variables: number density $n$, energy density $\cal E$, and the pressure $\cal P$ can be computed once the distribution function is specified. Below we list their definitions, respectively:
\ba
n &=& \intdP \, E f(x,p) \, , \\
{\cal E} &=& \intdP \, E^{2} f(x,p) \, , \\
{\cal P} &=& \frac{1}{3} \intdP \, p^{2} f(x,p) \, .
\ea
Here $dP$ is the Lorentz invariant momentum-space integration measure given by
\be
dP \equiv N_{\rm dof} \frac{d^3p}{(2\pi)^3} \frac{1}{E}  \, ,
\ee
where $N_{\rm dof}$ is the number of degrees of freedom. For compactness, $\tilde{N} \equiv N_{\rm dof}/(2\pi)^3$ is used below.

\subsection{Quasiparticle anisotropic hydrodynamics}
\label{subsec:qpah}

The quasiparticle Boltzmann equation is given by \cite{1509.02913,Jeon:1995zm,Romatschke:2011qp}
\be
p^\mu \partial_\mu f+\frac{1}{2}\partial_i m^2\partial^i_{(p)} f=-\mathcal{C}[f]\,,
\label{eq:boltz2}
\ee
where $p^{\mu}$ is the four momentum, $\partial_{\mu}$ is the four derivative, ($\mu$ for spatiotemporal indices and $i$ for spatial indices), $m$ is the mass, $f$ is the phase space density and the right hand side is the collision kernel containing all interactions. In this work, the collisional kernel is taken in the relaxation-time approximation (RTA), $\mathcal{C}[f]=p^\mu u_\mu(f-f_{\rm eq})/\tau_{\rm eq}$ with $\tau_{\rm eq}$ being the relaxation time~\cite{1509.02913}.

Taking the zeroth, first, and second moments of the Boltzmann equation yields, respectively, one obtains
\ba
\partial_\mu J^\mu&=&-\intdP \, {\cal C}[f]\, , \label{eq:J-conservation} \\
\partial_\mu T^{\mu\nu}&=&-\intdP \, p^\nu {\cal C}[f]\, , \label{eq:T-conservation} \\
\partial_\mu {\cal I}^{\mu\nu\lambda}- J^{(\nu} \partial^{\lambda)} m^2 &=&-\intdP \, p^\nu p^\lambda{\cal C}[f]\, \label{eq:I-conservation},
\ea 
where $J^\mu$ is the particle four-current, $T^{\mu\nu}$ is the energy-momentum tensor, and ${\cal I}^{\mu\nu\lambda}$ is a rank-three tensor. They are given by
\ba
J^\mu &\equiv& \intdP \, p^\mu f(x,p)\, , \label{eq:J-int} \\
T^{\mu\nu}&\equiv& \intdP \, p^\mu p^\nu f(x,p)+B g^{\mu\nu}, \label{eq:T-int}\\
{\cal I}^{\mu\nu\lambda} &\equiv& \intdP \, p^\mu p^\nu p^\lambda  f(x,p) \, .
\label{eq:I-int}
\ea
In Eq.~(\ref{eq:T-int}), a background contribution $B$ is introduced to ensure thermodynamic consistency as explained in detail in Ref.~\cite{1509.02913}. As a result, there exists a partial differential equation relating $B$ and the thermal mass
\be
\partial_\mu B = -\frac{1}{2} \partial_\mu m^2 \intdP  f(x,p)\,.
\label{eq:BM-matching}
\ee
The thermal mass $m(T)$ is obtained by tuning to the equation of state from lattice QCD calculations \cite{1007.2580}.  The procedure for performing this extraction is explained in detail in Ref.~\cite{1509.02913}.

\subsection{0+1D Quasiparticle anisotropic hydrodynamics}
\label{subsec:0p1qah}
In this work, we will limit ourselves to boost-invariant systems (0+1D). In this case, the energy density, transverse pressure, and longitudinal pressure are given by~\cite{1509.02913}
\ba
{\cal E} &=& \tilde{{\cal H}}_3({\boldsymbol\alpha},\hat{m}) \, \lambda^4+B \, ,\nonumber \\
{\cal P}_T &=& \tilde{{\cal H}}_{3T}({\boldsymbol\alpha},\hat{m}) \, \lambda^4-B \, ,\nonumber \\
{\cal P}_L &=& \tilde{{\cal H}}_{3L}({\boldsymbol\alpha},\hat{m}) \, \lambda^4-B \,.
\label{eq:E-P-trans}
\ea
In 0+1D,  Eq.~(\ref{eq:BM-matching}) can be written as
\be
\partial_\tau B = - \frac{\lambda^2 }{2} \tilde{\cal H}_{3B}({\boldsymbol\alpha},\hat{m}) \, \partial_\tau m^2  \, .
\label{eq:B}
\ee
For definitions of the various ${\cal{H}}$-functions appearing above, we refer the reader to App.~\ref{app:h-functions}. In the case where a Tsallis distribution is assumed, the $q$ dependence is implicit in $f(x,q)$, which the ${\cal{H}}$-functions are based upon.
%
\subsection{The dynamical equations in 0+1D}
\label{subsec:dyneq}
In the 0+1D case, there are five dynamical variables $\lambda$, $T$, $\alpha_x$, $\alpha_z$, and $q$. Hence we need 5 dynamical equations which can be obtained from the zeroth, first and second moments of the Boltzmann equation. 

From the zeroth moment, one obtains,
\be
\partial_\tau n+\frac{n}{\tau} =\frac{1}{\tau_{\rm eq}}\Big(n_{\rm eq}-n\Big)   \,,
\label{eq:othmoment}
\ee
where 
\be
n_{\rm eq} =4 \pi \tilde{N} T^3  \hat{m}_{\rm eq}^2 K_2\!\left( \hat{m}_{\rm eq}\right)\,.
\label{eq:neq1}
\ee
and
\be
n =4 \pi \tilde{N} \lambda^3  \, \alpha_x^2 \, \alpha_z \, \tilde{n}\!\left( \hat{m}\right)\, ,
\label{eq:neq2}
\ee
\be
\tilde{n}\!\left( \hat{m}\right) = \int d\hat{p} \, \hat{p}^2 f\!\left(\sqrt{\hat{p}^2 + \hat{m}^2} \right).
 \label{eq:Iiso1}
\ee
From the first moment, the conservation of energy-momentum $\partial_\mu T^{\mu\nu}=0$, one obtains,
\be
\partial_\tau {\cal E}=-\frac{{\cal E+P}_L}{\tau}\,.
\label{eq:1st-mom-u}
\ee
From the second moment,  Eq.~(\ref{eq:I-conservation}), one obtains
\ba
\partial_\tau \log{\cal I}_x+\frac{1}{\tau} &=&\frac{1}{\tau_{\rm eq}}\Big(\frac{{\cal I}_{\rm eq}}{{\cal I}_x}-1\Big) ,\label{eq:xx-trans}\\
\partial_\tau  \log{\cal I}_z+\frac{3}{\tau} &=&\frac{1}{\tau_{\rm eq}}\Big(\frac{{\cal I}_{\rm eq}}{{\cal I}_z}-1\Big) , \label{eq:zz-trans}
\ea
where
\ba
{\cal I}_{\rm eq}&=& 4 \pi   \tilde{N} T^5 \hat{m}_{\rm eq}^3 K_3\left( \hat{m}_{\rm eq}\right),  \\
{\cal I}_x&=& \frac{4 \pi}{3} \tilde{N} \lambda^5 \, \alpha_x^4 \, \alpha_z \, {\cal I}\left( \hat{m}\right),\label{eq:Ix}\\
{\cal I}_z &=& \frac{4 \pi}{3} \tilde{N} \lambda^5 \, \alpha_x^2 \, \alpha_z^3 \, {\cal I} \left( \hat{m}\right)\, . \label{eq:Iz}
\ea
with $K_3$ is a modified Bessel function of the second kind and ${\cal I}$ is given by
\be
{\cal I}\left( \hat{m}\right) = \int d\hat{p} \, \hat{p}^4 f\left(\sqrt{\hat{p}^2 + \hat{m}^2} \right).
 \label{eq:Iiso}
\ee
When $f(x)$ is in the Tsallis form, the integral in Eq.~(\ref{eq:Iiso}) can be done numerically,  where, upon using the Boltzmann distribution function, the following result is obtained
\be
{\cal I} (\hat{m})=3 \hat{m}^3 K_3\left( \hat{m}\right).
 \label{eq:}
\ee

The fifth equation necessary can be obtained by using the matching condition which reflects energy-momentum conservation: 
\be
\tilde{\cal H}_3 \lambda^4 = \tilde{\cal H}_{3,\rm eq} T^4.
 \label{eq:matching1}
\ee

By expanding Eqs.~(\ref{eq:othmoment}), (\ref{eq:1st-mom-u}), (\ref{eq:xx-trans}), (\ref{eq:zz-trans}) and (\ref{eq:matching1}), the final equations can be written in the following format: 
\ba
&&2\partial_\tau\log\alpha_x+\partial_\tau\log\alpha_z+3\partial_\tau\log\lambda
+ \frac{1}{\tilde{n}(\hat{m})} \tilde{n}'(\hat{m}) \partial_\tau\log \hat{m}
+ \frac{1}{\tilde{n}(\hat{m})} \tilde{n}'_q(\hat{m}) \partial_\tau q +\frac{1}{\tau} 
\nonumber \\ && \hspace{8cm} 
= \frac{1}{\tau_{\rm eq}}\left[\frac{1}{\alpha_x^2\alpha_z}\Big(\frac{T}{\lambda}\Big)^3  \frac{ \hat{m}_{\rm eq}^2 K_2(\hat{m}_{\rm eq})}{\tilde{n}(\hat{m})}-1\right] \,, 
\label{eq:final0th} 
\\
&& 4 \tilde{\cal H}_3 \partial_\tau\log\lambda+\tilde{\Omega}_m\partial_\tau\log \hat{m} +\tilde{\Omega}_L\partial_\tau\log\alpha_z
+\tilde{\Omega}_T\partial_\tau\log\alpha_x^2 + \tilde{\cal H}_{3q} \, \partial_\tau q+\frac{\partial_\tau B}{\lambda^4}+\frac{\tilde{\Omega}_L}{\tau} = 0 \, , \hspace{1cm}
 \label{eq:final1st} \\
&& 4\partial_\tau\log\alpha_x+\partial_\tau\log\alpha_z+5\partial_\tau\log\lambda
+ \frac{1}{{\cal I}(\hat{m})} {\cal I}'(\hat{m}) \partial_\tau\log \hat{m}
+ \frac{1}{{\cal I}(\hat{m})} {\cal I}'_q(\hat{m}) \partial_\tau q +\frac{1}{\tau} 
\nonumber \\ && \hspace{8cm} 
= \frac{1}{\tau_{\rm eq}}\left[\frac{1}{\frac{1}{3}\alpha_x^4\alpha_z}\Big(\frac{T}{\lambda}\Big)^5  \frac{ \hat{m}_{\rm eq}^3 K_3(\hat{m}_{\rm eq})}{{\cal I}(\hat{m})}-1\right] \,, 
\label{eq:final2nd1} 
\\
&& 2\partial_\tau\log\alpha_x+3\partial_\tau\log\alpha_z+5\partial_\tau\log\lambda
+ \frac{1}{{\cal I}(\hat{m})} {\cal I}'(\hat{m}) \partial_\tau\log \hat{m}
+ \frac{1}{{\cal I}(\hat{m})} {\cal I}'_q(\hat{m}) \partial_\tau q +\frac{3}{\tau}
 \nonumber \\ && \hspace{8cm} 
= \frac{1}{\tau_{\rm eq}}\left[\frac{1}{\frac{1}{3}\alpha_x^2\alpha_z^3}\Big(\frac{T}{\lambda}\Big)^5 \, \frac{ \hat{m}_{\rm eq}^3 K_3(\hat{m}_{\rm eq})}{{\cal I}(\hat{m})}-1\right] \,,
\label{eq:final2nd2}
\ea
\be
4 \tilde{\cal H}_{3,\rm eq} \partial_\tau\log T +\tilde{\Omega}_{m,\rm eq}\partial_\tau\log \hat{m}_{\rm eq}
+\frac{\tilde{\Omega}_L}{\tau}\Big(\frac{\lambda}{T}\Big)^4+\frac{\partial_\tau B}{T^4}=0\,,
\label{eq:final-matching}
\ee
where $\tau_{\rm eq}$ is a function of $T(\tau)$ and the temperature-dependent quasiparticle mass $m(T)$~\cite{1509.02913}.   When using a Boltzmann distribution, there are four dynamical equations obtained solely from the first and second moments; derived in detail in Ref.~\cite{1509.02913}. They also can be obtained from Eqs.~(\ref{eq:final1st})-(\ref{eq:final-matching}) by setting $\partial_\tau q=0$ and using the Boltzmann distribution in the definitions of the $\tilde{\cal H}$, $\Omega$, and ${\cal I}$ functions. The reader may refer to App.~\ref{app:h-functions} for more details about the special functions specified above.

One should note that, from the definitions of the bulk variables, one needs to obtain $B$ in order to obtain the full energy density and pressures. For this purpose, we integrate the dynamical equations to very late times ($\tau_f=100$ fm/c) and then integrate Eq.~(\ref{eq:B}) backwards in time using $B(\tau_f)=B_{\rm eq}(T(\tau_f))$ since $B_{\rm eq}(T=0)=0$~\cite{1509.02913}.

\begin{figure}[t!]
\centerline{
\hspace{-1.5mm}
\includegraphics[width=1\linewidth]{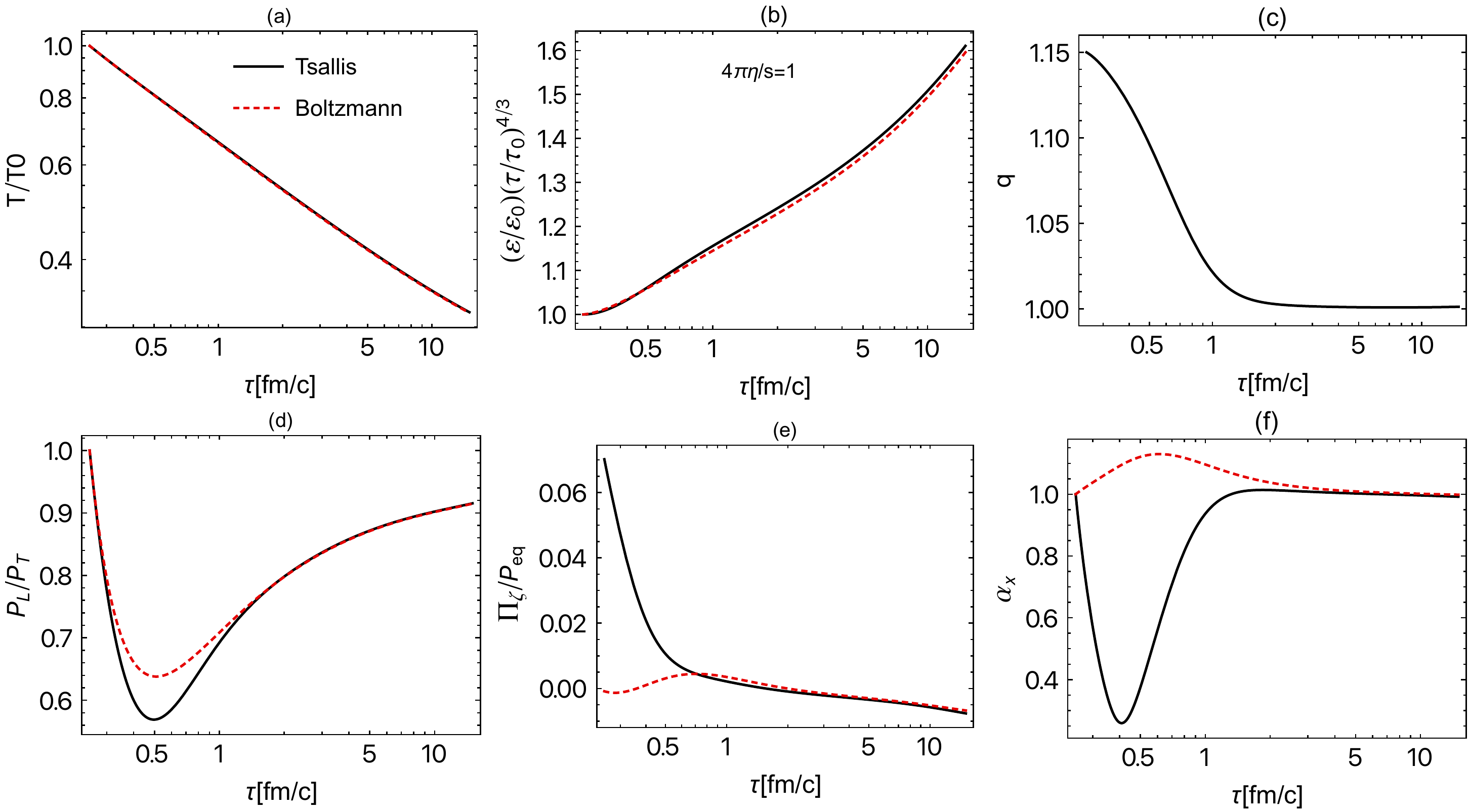}}
\caption{Top row: proper-time evolution of scaled effective temperature,  scaled energy density, and the Tsallis parameter. Bottom row: proper-time evolution of  the pressure anisotropy, bulk pressure, and the anisotropy parameter $\alpha_x$. The black solid line represents results using a Tsallis distribution with initial deviation parameter $\delta q=0.15$, whereas the red dashed line represents results using a Boltzmann  distribution. In this figure, the shear viscosity to entropy density ratio used is $4 \pi\eta/s=1$.}
\label{fig:bulkVar1}
\end{figure}

\begin{figure}[t!]
\centerline{
\hspace{-1.5mm}
\includegraphics[width=1\linewidth]{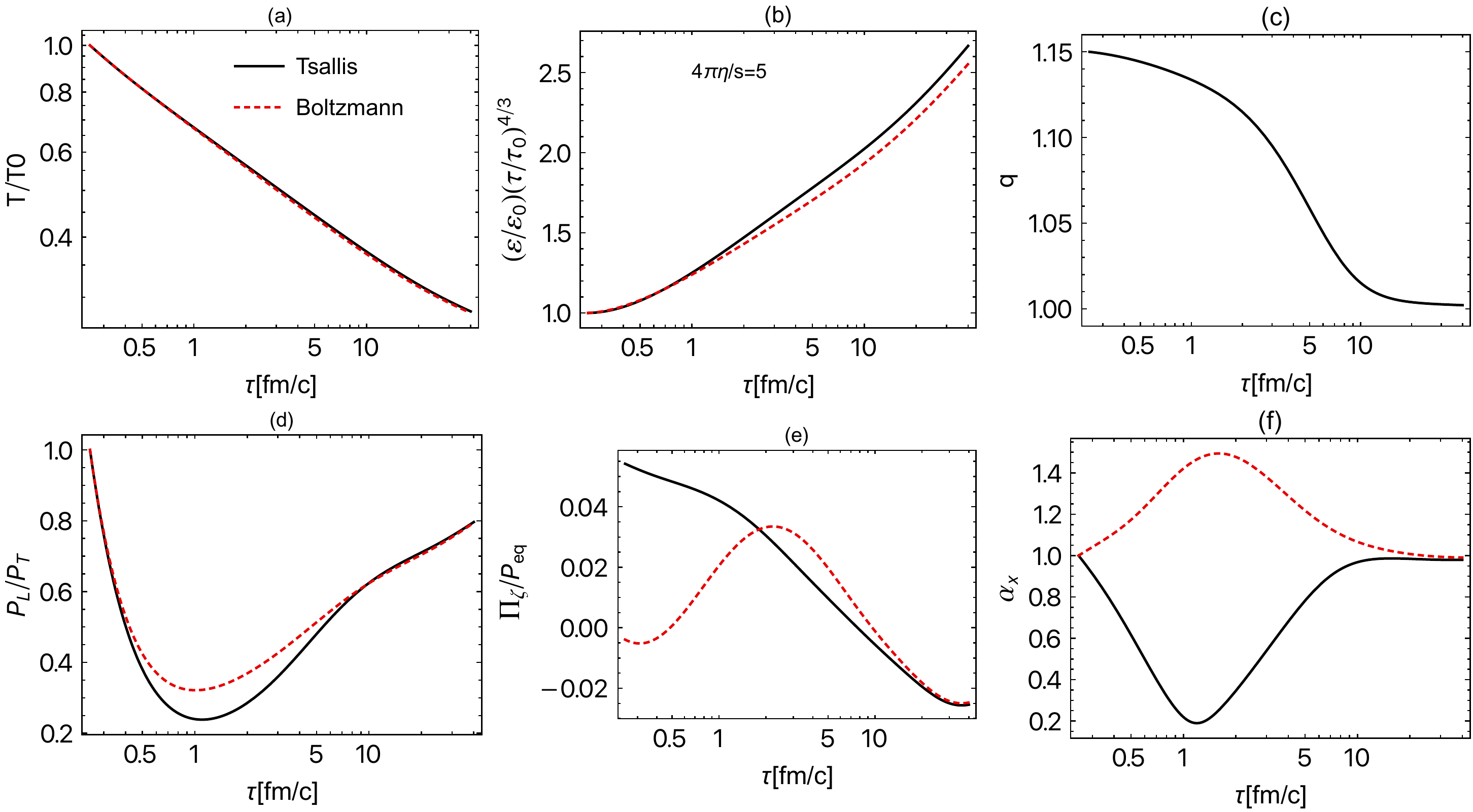}}
\caption{Same as Fig.~\ref{fig:bulkVar1} with shear viscosity to entropy density ratio $4 \pi\eta/s=5$.}
\label{fig:bulkVar5}
\end{figure}

\section{Results}
\label{sec:results}

In this section, we present comparisons of the 0+1D quasiparticle anisotropic hydrodynamics model using both {the} Boltzmann and Tsallis distribution functions. For reference, we compare the results of this work to the results presented in Ref.~\cite{1509.02913}, where a Boltzmann distribution function is used instead of the Tsallis  distribution function. We first solve the dynamical equations presented in the last section subject to this set of initial conditions: $T_0(\tau_0)=600$ MeV, $\alpha_x(\tau_0)=1$, and $\alpha_z(\tau_0)=1$ at an initial proper time  $\tau_0=0.25$ fm/c. Then we show differences in the temporal evolution of temperature, energy density, pressure anisotropy, and bulk pressure, which is defined as
\be
\Pi \equiv \frac{1}{3} \left({\cal{P}}_{\rm{L}}+ 2 {\cal{P}}_{\rm{T}}\right)-{\cal{P}}_{\rm{eq}} \, .
 \label{eq:bulkpress_defn}
\ee

\subsection{Proper-time evolution}

In Figs.~\ref{fig:bulkVar1}-\ref{fig:bulkVar5}, we use black solid and red dashed lines for results of the Tsallis and Boltzmann distribution functions, respectively. In the Tsallis approach, we also assumed an initial Tsallis parameter $q (\tau_0)= 1.15$.  In the top row of Fig.~\ref{fig:bulkVar1}, we show the proper-time evolution of the scaled effective temperature, scaled energy density, and the Tsallis parameter. In the bottom row of Fig.~\ref{fig:bulkVar1}, we show the proper-time evolution of the pressure anisotropy, bulk pressure, and the transverse anisotropy parameter $\alpha_x$. In this figure, the shear viscosity to entropy density ratio is taken to be $4 \pi\eta/s=1$. As can be seen from the figure, the effective temperature evolution in both approaches is identical whereas some differences are seen in the evolution of the scaled energy density and pressure anisotropy. The maximum difference for the scaled energy density is roughly $\sim 1 $\% at late times whereas the maximum difference for the pressure anisotropy is roughly $\sim 10$\% at $\tau \sim 0.5$ fm/c. The temporal evolution of the bulk pressure shows clear differences between the two methods which means the bulk pressure is sensitive to which distribution function is assumed. We note that differences in the bulk evolution exist between these two approaches even for very small initial $q$ values such as $\delta q= 0.001$ (not shown here). As shown in Ref.~\cite{1605.02101}, differences in the evolution of the bulk pressure can have a direct impact on the primordial particle spectra and hence this may be important phenomenologically. Next, we show that, at late times, both the Tsallis parameter $q$ and the transverse anisotropy parameter $\alpha_x$ approach unity as expected by using the RTA approximation. 
 
In Fig.~\ref{fig:bulkVar5}, we show the proper-time evolution of the same quantities as in Fig.~\ref{fig:bulkVar1} with the shear viscosity to entropy density ratio is taken to be $4 \pi\eta/s=5$. As can be seen from this figure, deviations between the two approaches become more pronounced when increasing the shear viscosity to entropy density ratio. We note, as can be seen from panels (e) and (f), that it takes a longer time for both approaches to converge to a universal results, compared to results in Fig.~\ref{fig:bulkVar1} for the same panels.

We note that in the Tsallis approach, there are numerical instabilities appearing at late times resulting from integrals that depend on derivatives of $f_T$ as shown in App.~\ref{app:h-functions}. Although they do not affect the late time behavior of some quantities such as the pressure anisotropy, they affect late time behavior of the bulk pressure. To control this numerical issue, we take two steps. First, we expand $f_T(x,q)$ around $q\rightarrow 1$ up to the sixth-order and use this approximated $f_T$ when $q$ is close to $1$, this point is taken to be $q_c=1.00001$. Second, since the integrands are practically zero at large $x$, we put a limit on the integrations instead of integrating to $\infty$. With these changes, we still see some numerical issues at late times, however, we always terminate evolution before they become too serious.
 
\begin{figure}[t!]
\centerline{
\includegraphics[width=0.99\linewidth]{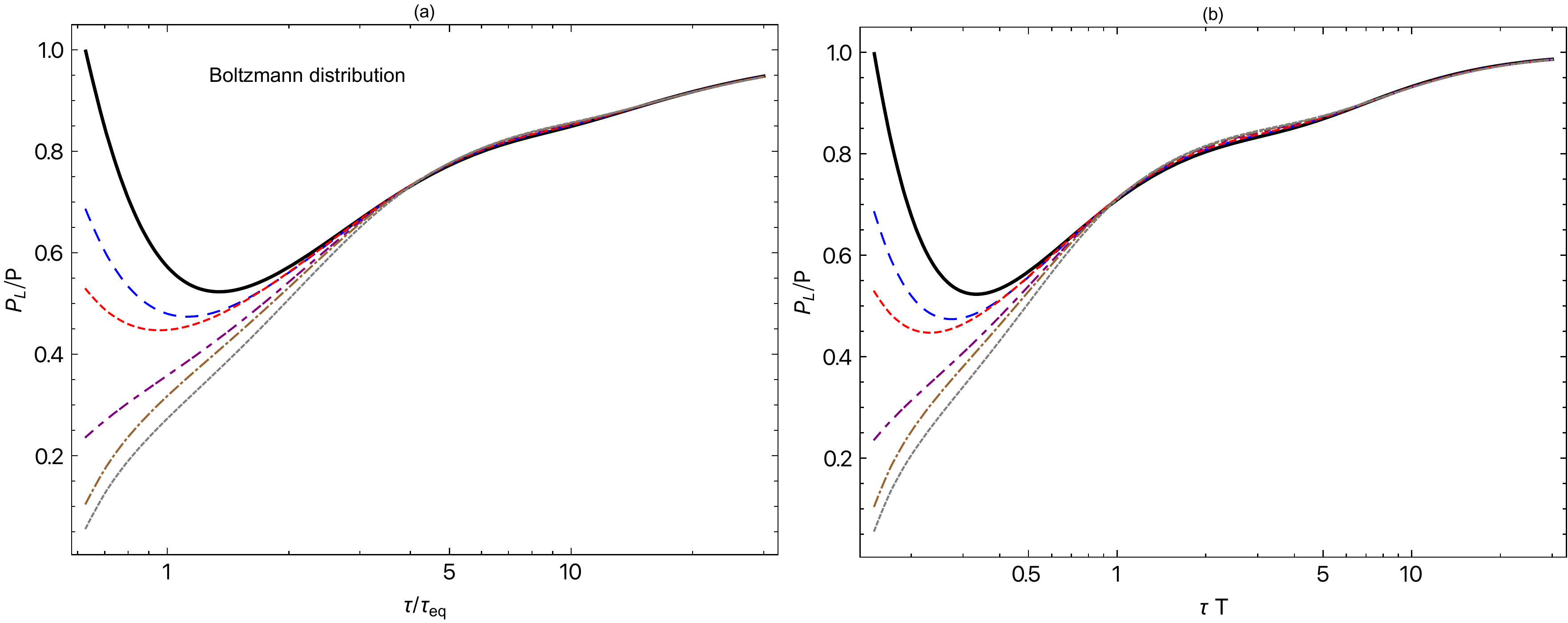}}
\centerline{
\includegraphics[width=0.99\linewidth]{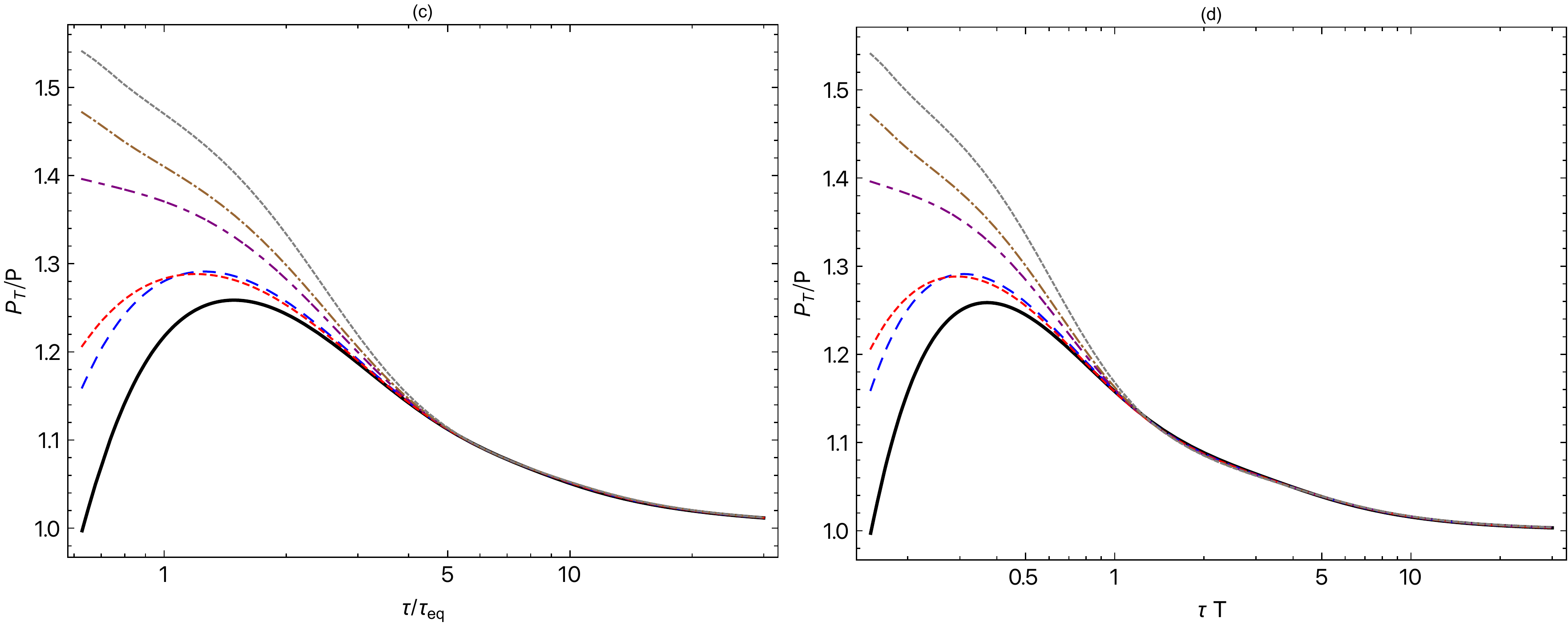}}
\centerline{
\includegraphics[width=0.99\linewidth]{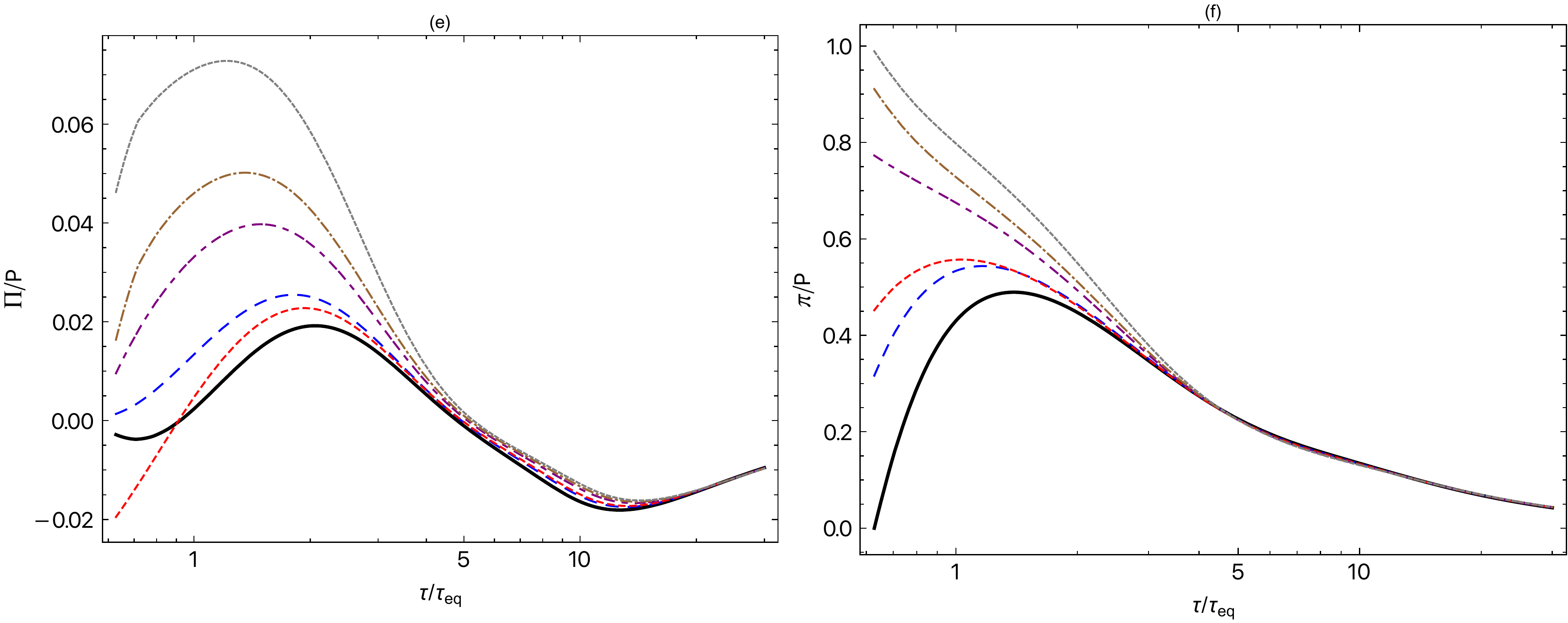}
}
\caption{Top row: Evolution of $P_L/P_{\rm eq}$ as a function of $\tau/\tau_{\rm eq}$ and $\tau T$, in the left and right panels, respectively. Middle row: Evolution of $P_T/P_{\rm eq}$ as a function of $\tau/\tau_{\rm eq}$ and $\tau T$, in the left and right panels, respectively. Bottom row: Evolution of $\Pi/P_{\rm eq}$ and $\pi/P_{\rm eq}$ as a function of $\tau/\tau_{\rm eq}$, in the left and right panels, respectively.}
\label{fig:BoltzmannAttractor}
\end{figure}

\subsection{Non-equilibrium attractors}

We next turn to a discussion of non-equilibrium attractors.  For this purpose we will make plots of the shear viscous stress, which is defined as
\be
\pi \equiv \frac{2}{3} \left({\cal{P}}_{\rm{T}}- {\cal{P}}_{\rm{L}}\right) \, .
 \label{eq:shearpress_defn}
\ee

\begin{figure}[t!]
\centerline{
\includegraphics[width=0.99\linewidth]{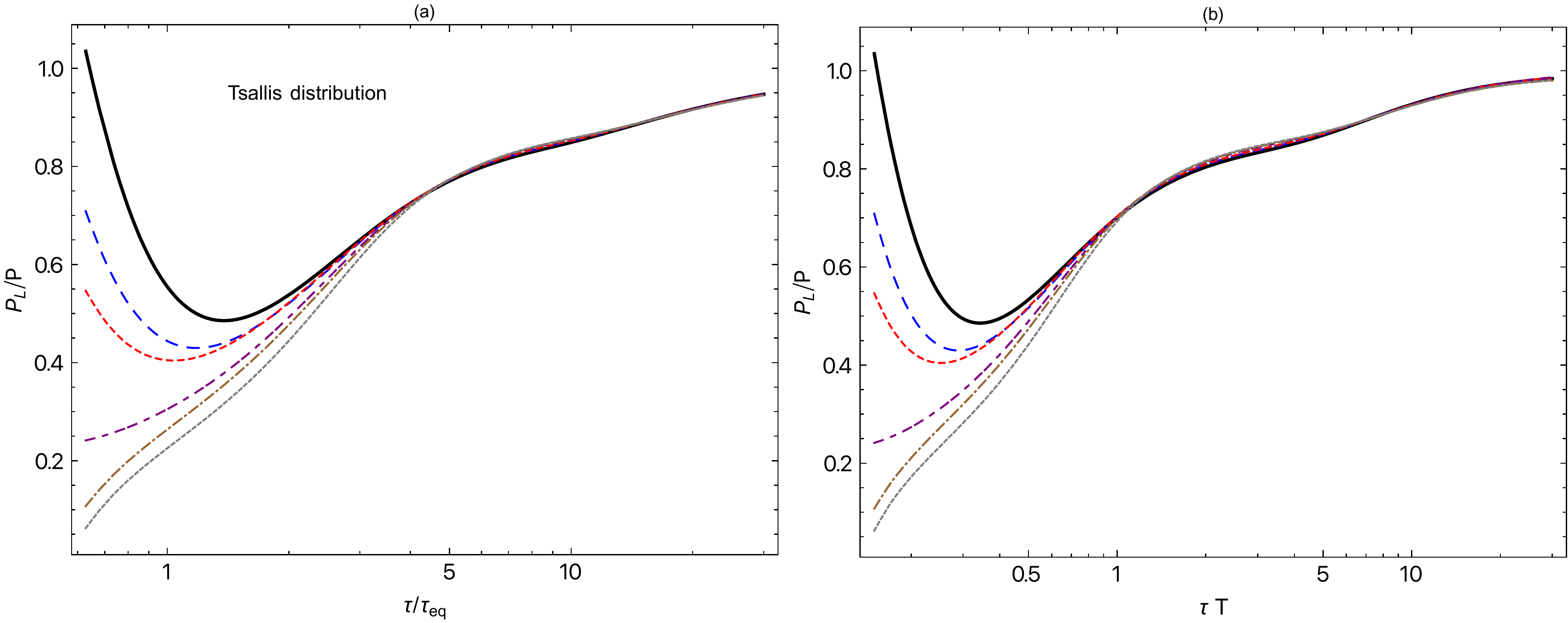}}
\centerline{
\includegraphics[width=0.99\linewidth]{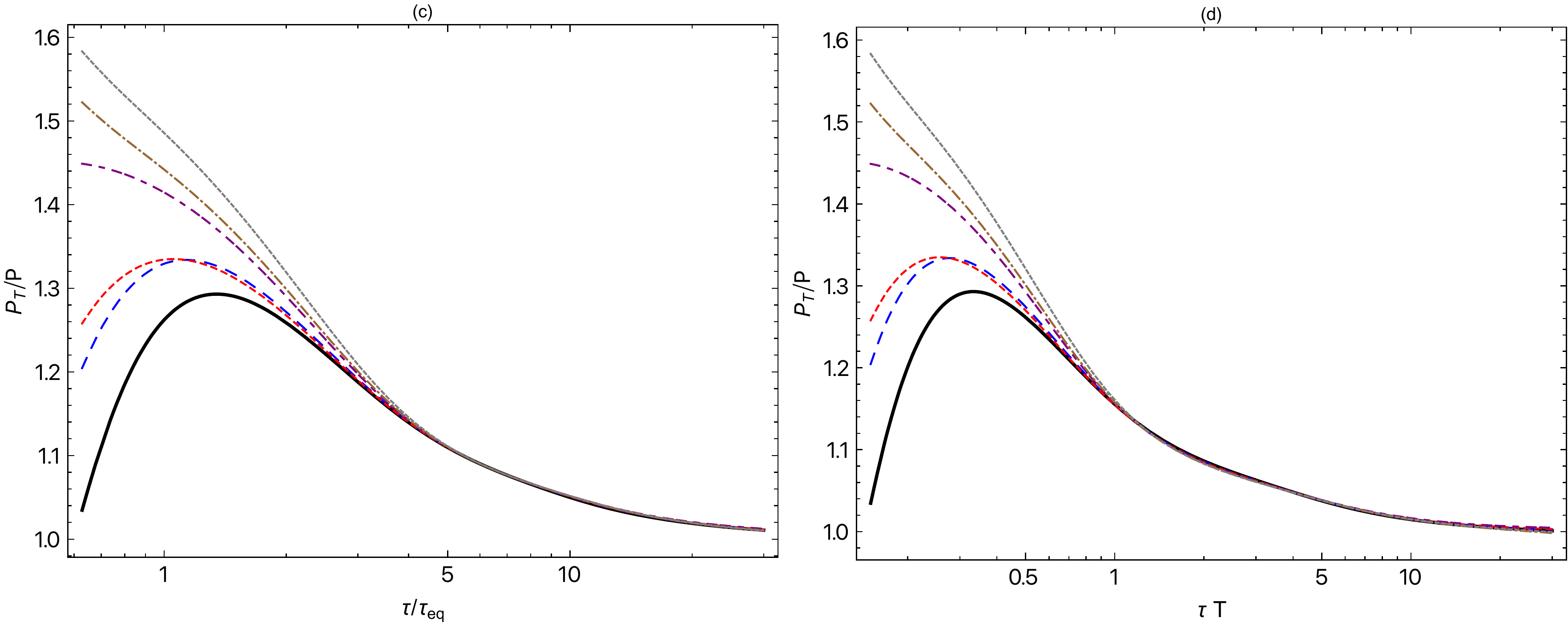}}
\centerline{
\hspace{-1.5mm}
\includegraphics[width=0.99\linewidth]{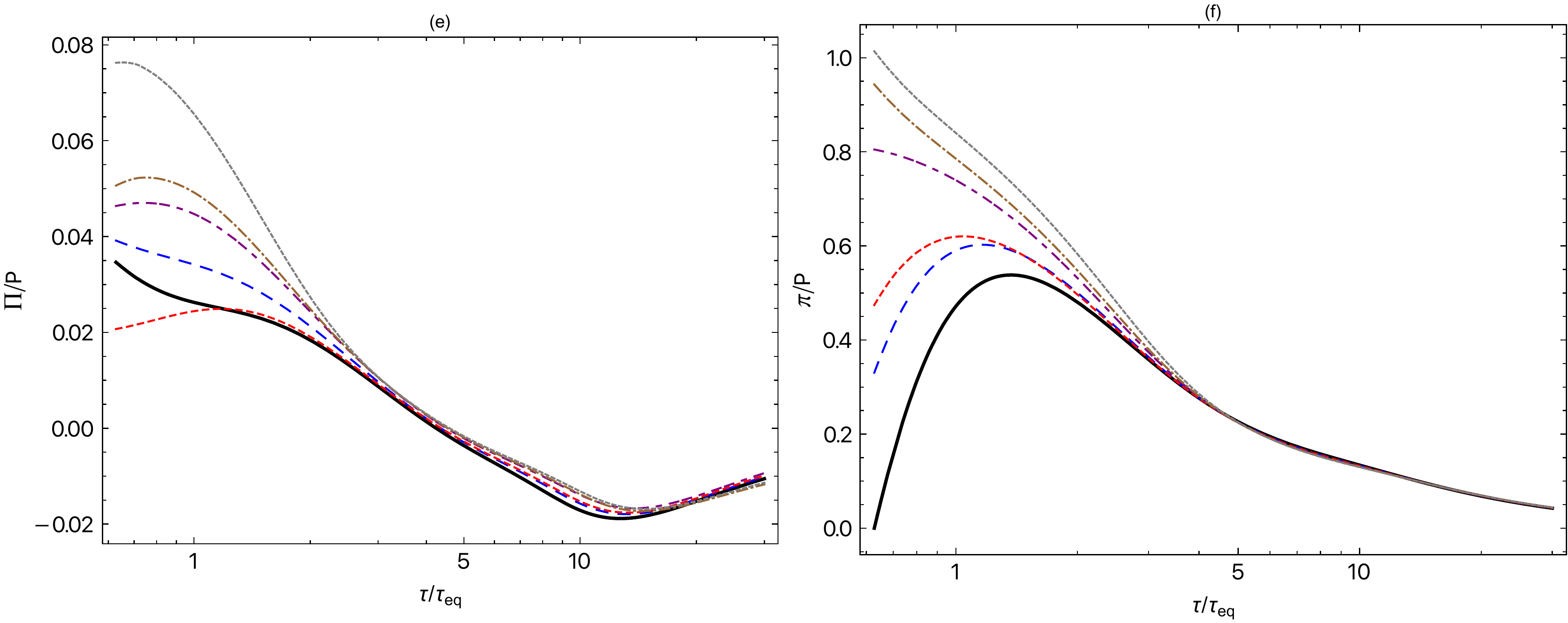}
}
\caption{Same as Fig.~\ref{fig:BoltzmannAttractor} with Tsallis distribution used instead of Boltzmann distribution. Similar initial conditions are used as in  Fig.~\ref{fig:BoltzmannAttractor} with $q_0$ taken to be  $1.1$ in all panels and for all curves. }
\label{fig:TsallisAttractor}
\end{figure}

In Figs.~\ref{fig:BoltzmannAttractor}-\ref{fig:TsallisAttractor}, we investigate the attractor behavior corresponding to a variety of initial conditions using Boltzmann and Tsallis distributions, respectively.   In both figures, we assumed $4 \pi\eta/s=3$ and $\delta q (\tau_0)= 0.10$. We show the scaled time evolution $\tau/\tau_{\rm eq}$ and for reference, in some cases, we also show $\tau T$ evolution which is the limit for the conformal relaxation time, i.e.  $\tau_{\rm eq} \sim 1/T(\tau)$. We note that the black-solid line is initially isotropic, i.e. $\alpha_x=\alpha_z=1$. As a result, $P_L/P_{\rm eq}=P_T/P_{\rm eq}=1$, whereas the bulk and shear viscous stress are both equal to zero. In the same figures, the other colored lines correspond to different initial conditions where $\alpha_x \neq \alpha_z \neq 1$. Finally, to put the numbers presented in perspective, we note that at $\tau_0$,  $\tau T=0.15$ and $\tau/\tau_{\rm eq} = 0.628217$ and, at very late times, e.g. $\tau =100$ fm/c,  $\tau T \sim 15$ and $\tau/\tau_{\rm eq} \sim 35$.
 
In Fig.~\ref{fig:BoltzmannAttractor}, top and middle rows, we show the evolution of the scaled longitudinal pressure $P_L/P_{\rm eq}$ and the scaled transverse pressure $P_T/P_{\rm eq}$, respectively. Both $P_L/P_{\rm eq}$ and $P_T/P_{\rm eq}$ are shown as a function of $\tau/\tau_{\rm eq}$ and $\tau T$ in the left and right panels, respectively. We see that all results for different set of initial conditions converge to a universal curve at early times $\tau \sim 4 $ fm/c. In the bottom row, we plot results of the scaled bulk pressure $\Pi/P_{\rm eq}$ and the shear stress $\pi/P_{\rm eq}$ as a function of $\tau/\tau_{\rm eq}$. We note that results for $\pi/P_{\rm eq}$ approach a universal attractor at an earlier rescaled time $\tau/\tau_{\rm eq} \sim 4$ while the  $\Pi/P_{\rm eq}$ converges later, only for $\tau/\tau_{\rm eq} > 10$ due to the system reaching isotropic equilibrium. In the context of initial conditions used here $\tau/\tau_{\rm eq} =4$ corresponds to $\tau \sim 4 $ fm/c, whereas $\tau/\tau_{\rm eq} =10$ corresponds to $\tau \sim 20$ fm/c.
 
In Fig.~\ref{fig:TsallisAttractor}, we use a Tsallis distribution and present the same plots as shown in Fig.~\ref{fig:BoltzmannAttractor}. From all panels, we see the existence of a universal attractor even for this nonextensive far-from equilibrium approach. The late time differences in the $\Pi/P_{\rm eq}$ evolution seen in panel (e) could be purely numerical in origin; however, the differences seen at earlier rescaled times indicate that, strictly speaking, there does not exist a hydrodynamic attractor for the evolution of the bulk viscous pressure.  Despite this, similar to as was found using exact solutions in the relaxation time approximation~\cite{2107.10248}, we find that the transverse and longitudinal pressure ratios shown in panels (a) and (c) and the shear correction shown in panel (f) suggest the existence of  a non-equilibrium hydrodynamic attractor even when the system has a realistic non-conformal equation of state.  This is true for both Boltzmann and Tsallis statistics.
 
\begin{figure}[t!]
\centerline{
\hspace{-1.5mm}
\includegraphics[width=1\linewidth]{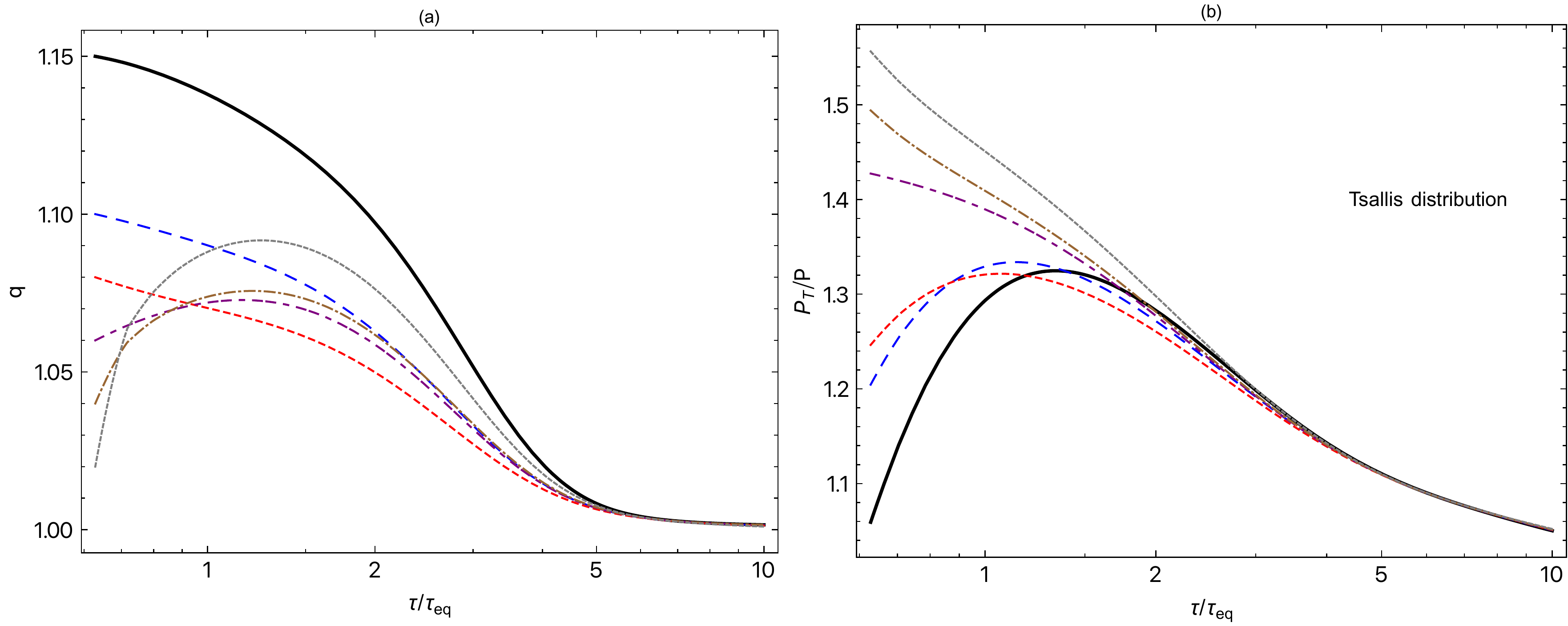}}
\caption{Evolution of $q$ and $P_T/P$ as a function of scaled timed $\tau/\tau_{\rm eq}$, left and right panels, respectively. Note that $q_0$ is different in each curve as can be seen from left panel whereas other initial conditions are the same as the ones used in  Fig.~\ref{fig:TsallisAttractor}.}
\label{fig:qPTP}
\end{figure}

Finally, in Fig.~\ref{fig:qPTP}, we change the initial condition for the Tsallis parameter, $q_0$, for each curve, with the other initial conditions being the same as in Fig.~\ref{fig:TsallisAttractor}. In the left panel, we show the scaled time evolution of the Tsallis parameter $q$. As can be seen from this figure, all results converges quickly to unity at $\tau/\tau_{\rm eq} \sim 5$. In the right panel, as a cross check, we plot the scaled time evolution of $P_T/P_{\rm eq}$ for this set of initial condition and we can see that a universal attractor still exists. Figure~\ref{fig:qPTP} demonstrates that an attractor exists for the non-extensivity parameter $q$, with $q$ approaching unity from above.  This is a non-trivial finding and suggests that one can constrain the late-time value of $q$ used in phenomenological applications without a full dynamical simulation.

We note that fits to the experimental data indicate that the fitted Tsallis parameter $q$ is always close to unity meaning $\delta q$ is small. For example,  in Au+Au collisions at 200 GeV, fits to the spectra result in $\delta q = 0.015$ and $\delta q =0.086$ at 10-20\% and 60-80\%, respectively~\cite{0812.1609}. In Pb-Pb collisions at 2.76 TeV, fits to the spectra result in $\delta q = 0.1363$ at 5-10 \% centrality~\cite{1911.04878}. Moreover, in p-p collisions at 5.02 TeV and 13 TeV, fits to the spectra data result in $\delta q \sim 0.12$ which increases as a function of multiplicity to reach  $\delta q \sim 0.14$ ~\cite{1908.04208}. For results presented in this work, we used different values of $\delta q\in \{0.15, 0.10, 0.08, 0.06, 0.04, 0.02\}$. In practice, one would like to go to higher values of $\delta q$; however, for large $\delta q$ the moment expansion of the distribution is ill-defined due to UV divergences. In addition, we find that our numerics become quite unstable for large $\delta q$ close to this limit.  This is expected since the $\cal{H}$ functions introduced in the App.~\ref{app:h-functions}, which are part of the dynamical equations, each have their own convergence intervals. The convergence intervals for the bulk variables in rather simplified systems are shown in Ref.~\cite{1608.08965}. As an example, in the massless limit, the number density is convergent only in the interval $0 \leq q \leq 1.5$.


\section{Conclusions and Outlook}
\label{sec:conclusions}

In this paper, we reviewed the basics of quasiparticle anisotropic hydrodynamics which is based on the self-consistent introduction of a single temperature-dependent quasiparticle mass for the degrees of freedom in the Boltzmann equation.  The temperature dependence of the quasiparticle mass was determined uniquely by matching to lattice QCD calculations of the equation of state. We then used 0+1D quasiparticle anisotropic hydrodynamics as a dynamical model, with the underlying distribution given by either a Tsallis distribution function or its $\delta q \rightarrow 0$ limit, which corresponds to a Boltzmann distribution function. We then compared the temporal evolution of temperature, pressure anisotropy, energy density, and bulk pressure predicted by the two approaches at different shear viscosity to entropy density ratios. This work demonstrates that the temperature, pressure anisotropy, and energy density evolutions are not very sensitive to which distribution function is used in the model. However, we found that the bulk evolution is sensitive to which distribution function is assumed even for very small $\delta q \sim 0.001$. 

In the last section of the paper we demonstrated the existence of hydrodynamic attractors in both non-conformal cases. We found that early-time hydrodynamic attractors exist for the scaled longitudinal and transverse pressures even though the system is non-conformal and non-extensive. We also found that, although the bulk viscous pressure does not have an early-time hydrodynamic attractor, the shear stress seems to converge to an early-time hydrodynamic attractor.  These results are in agreement with earlier findings which made use of exact solutions to the 0+1D RTA Boltzmann equation with Boltzmann statistics~\cite{2107.10248}.  Finally, by varying $q_0$, we found that there exists a hydrodynamic attractor for the Tsallis parameter $q$ when $q$ is plotted as a function of rescaled time.  This observation can help to strongly constrain late-time values of $q$ for phenomenological applications.

 As shown in Refs.~\cite{1509.02913,1605.02101}, differences in the evolution of the bulk pressure can have a direct impact on the primordial hadron spectra. Looking forward, these differences could allow a determination of the optimal form of the distribution function on the freeze-out hypersurface. This could be done by modifying the existing 3+1d aHydroQP code where the distribution function is assumed to be in the Boltzmann form~\cite{Alqahtani:2017jwl}, to study the effect of Tsallis statistics on heavy-ion observables such as the spectra especially at intermediate $p_T\sim3$ GeV. On the theory front, it would be quite interesting to study the conformal hydrodynamic attractor in the case of Tsallis distribution at both zero and finite chemical potential as performed in Refs.~\cite{Strickland:2017kux,Almaalol:2018ynx,Almaalol:2018jmz}. Additionally, one also may look for the existence of the exact solutions to the Boltzmann equation using a Tsallis distribution function at small $\delta q$~\cite{Florkowski:2013lya}.  These projects are planned for future follow-up work. 
 

\acknowledgments{ }
M. Alqahtani was supported by the Deanship of Scientific Research at the Imam Abdulrahman Bin Faisal University under grant number 2021-089-CED. M. Strickland was supported by the U.S. Department of Energy, Office of Science, Office of Nuclear Physics under Award No. DE-SC0013470.

\appendix 
\section{Special functions}
\label{app:h-functions}

For completeness,  we list below the special functions that appear in the main body of the manuscript. We note here that the only difference between the two approaches compared in this work is the distribution function $f(x)$ which is assumed to be in the Tsallis  form (the new approach) and the Boltzmann form, i.e., $f(x)={\rm exp}(-x)$ (most of which are presented before in ~Ref.~\cite{1509.02913}).

In the Tsallis form, the derivative with respect to the argument, $x$ gives 
\be
f'(x)=-[1+ \delta q \, x]^{-\frac{q}{\delta q}}\,,
\label{eq:TsallisDist3}
\ee
and the derivative with respect to $q$ gives
\be
f'_q(x) \equiv \partial_q f(x)=\left(1+ \delta q \, x\right)^{-\frac{1}{\delta q}} \left(-\frac{x}{\delta q (1+\delta q\, x)} + \frac{\log(1+\delta q \, x)}{\delta q^2}          \right) .
\label{eq:TsallisDist4}
\ee

\subsection{Zeroth moment}

For the evaluation of the zeroth moment of the Boltzmann equation we need
\be
\tilde{n}'(\hat{m}) \equiv  \hat{m} \, \partial_{\hat{m}}\tilde{n} =\hat{m}^2
\int_0^\infty d\hat{p} \, \hat{p}^2  \frac{f' \!\left(\!\sqrt{\hat{p}^2 + \hat{m}^2}\right)}{\sqrt{\hat{p}^2 + \hat{m}^2}} \, ,
\label{eq:nptilde}
\ee
where we have multiplied by $\hat{m}$ to keep $ \partial_\tau\log \hat{m}$ similar to other terms. 

Additionally, one has
\be
\tilde{n}'_q (\hat{m}) \equiv \partial_{q}\tilde{n}= \int d\hat{p} \,  \hat{p}^2  f'_q \!\left(\!\sqrt{\hat{p}^2 + \hat{m}^2}\right).
 \label{eq:matching}
\ee
\subsection{First moment}

The following special functions are related to the first moment of the Boltzmann equation
\ba
\tilde{\cal H}_3({\boldsymbol\alpha},\hat{m}) &\equiv&  2 \pi \tilde{N} \alpha_x^4
\int_0^\infty d\hat{p} \, \hat{p}^3  f\!\left(\!\sqrt{\hat{p}^2 + \hat{m}^2}\right) {\cal H}_2\!\left(\frac{\alpha_z}{\alpha_x},\frac{\hat{m}}{\alpha_x\hat{p}} \right) ,
\label{eq:h3tilde}
\\
\tilde{\cal H}_{3T}({\boldsymbol\alpha},\hat{m}) &\equiv&  \pi \tilde{N} \alpha_x^4
\int_0^\infty d\hat{p} \, \hat{p}^3  f\!\left(\!\sqrt{\hat{p}^2 + \hat{m}^2}\right) {\cal H}_{2T}\!\left(\frac{\alpha_z}{\alpha_x},\frac{\hat{m}}{\alpha_x\hat{p}} \right) ,
\label{eq:h3ttilde}
\\
\tilde{\cal H}_{3L}({\boldsymbol\alpha},\hat{m}) &\equiv&  2 \pi \tilde{N} \alpha_x^4
\int_0^\infty d\hat{p} \, \hat{p}^3  f\!\left(\!\sqrt{\hat{p}^2 + \hat{m}^2}\right) {\cal H}_{2L}\!\left(\frac{\alpha_z}{\alpha_x},\frac{\hat{m}}{\alpha_x\hat{p}} \right) ,
\label{eq:h3ltilde}
\\
\tilde{\cal H}_{3m}({\boldsymbol\alpha},\hat{m}) &\equiv&  -2 \pi \tilde{N} \alpha_x^4 \hat{m}^2
\int_0^\infty d\hat{p} \, \hat{p}^3  \frac{f' \!\left(\!\sqrt{\hat{p}^2 + \hat{m}^2}\right)}{\sqrt{\hat{p}^2 + \hat{m}^2}} {\cal H}_2\!\left(\frac{\alpha_z}{\alpha_x},\frac{\hat{m}}{\alpha_x\hat{p}} \right) ,
\label{eq:h3mtilde}
\\
\tilde{\cal H}_{3B}({\boldsymbol\alpha},\hat{m}) &\equiv &  2\pi \tilde{N}\alpha_x^2
\int_0^\infty d\hat{p} \, \hat{p} f\!\left(\!\sqrt{\hat{p}^2 + \hat{m}^2}\right) {\cal H}_{\rm 2B}\!\left(\frac{\alpha_z}{\alpha_x},\frac{\hat{m}}{\alpha_x \hat{p}} \right) .
\label{eq:h3Btilde} 
\ea
including a new integral of the form
\be
\tilde{\cal H}_{3q}({\boldsymbol\alpha},\hat{m}) \equiv  2 \pi \tilde{N} \alpha_x^4
\int_0^\infty d\hat{p} \, \hat{p}^3  f'_q\!\left(\!\sqrt{\hat{p}^2 + \hat{m}^2}\right) {\cal H}_2\!\left(\frac{\alpha_z}{\alpha_x},\frac{\hat{m}}{\alpha_x\hat{p}} \right).
\label{eq:h3qtilde}
\ee
The ${\cal H}_2$ functions are the same as the ones shown in App. B of~\cite{1509.02913}
\ba
 {\cal H}_2(y,z)
&=& \frac{y}{\sqrt{y^2-1}} \left[ (z^2+1)
\tanh^{-1} \sqrt{\frac{y^2-1}{y^2+z^2}} + \sqrt{(y^2-1)(y^2+z^2)} \, \right] ,\\
\label{eq:H2}
{\cal H}_{2T}(y,z) 
&=& \frac{y}{(y^2-1)^{3/2}}
\left[\left(z^2+2y^2-1\right) 
\tanh^{-1}\sqrt{\frac{y^2-1}{y^2+z^2}}
-\sqrt{(y^2-1)(y^2+z^2)} \right] , \hspace{1cm} \\
{\cal H}_{2L}(y,z)
&=& \frac{y^3}{(y^2-1)^{3/2}}
\left[
\sqrt{(y^2-1)(y^2+z^2)}-(z^2+1)
\tanh^{-1}\sqrt{\frac{y^2-1}{y^2+z^2}} \,\,\right],\\
\label{eq:H2L}
{\cal H}_{2B}(y,z)&\equiv& {\cal H}_{2T}(y,z)+ \frac{{\cal H}_{2L}(y,z)}{y^2}=\frac{2}{\sqrt{y^2-1}}\tanh^{-1} \sqrt{\frac{y^2-1}{y^2+z^2}} \, .
\ea
In addition, the $\Omega$ functions appearing in the dynamical equations are defined as linear combinations of ${\cal H}_3$ functions
\ba
\Omega_T({\boldsymbol\alpha},\hat{m}) &\equiv& {\cal H}_3+{\cal H}_{3T}\,,  \\
\Omega_L({\boldsymbol\alpha},\hat{m}) &\equiv& {\cal H}_3+{\cal H}_{3L}\,,  \\
\Omega_m({\boldsymbol\alpha},\hat{m}) &\equiv& {\cal H}_3-{\cal H}_{3L}-2{\cal H}_{3T}-{\cal H}_{3m}\,.
\ea

Finally, we note that in deriving the dynamical equations, the following identities are needed:
\ba
\frac{\partial {\cal H}_2(y,z)}{\partial y}&=&\frac{1}{y}\Big[{\cal H}_2(y,z)+{\cal H}_{2L}(y,z)\Big] , \\
\frac{\partial {\cal H}_2(y,z)}{\partial z}&=&\frac{1}{z}\Big[{\cal H}_2(y,z)-{\cal H}_{2L}(y,z)-{\cal H}_{2T}(y,z)\Big] ,
\ea
and
\ba
\frac{\partial \tilde{\cal H}_3}{\partial\alpha_x}&=&\frac{2}{\alpha_x}\tilde\Omega_T\,, \\
\frac{\partial \tilde{\cal H}_3}{\partial\alpha_z}&=&\frac{1}{\alpha_z}\tilde\Omega_L\,, \\
\frac{\partial \tilde{\cal H}_3}{\partial\hat{m}}&=&\frac{1}{\hat{m}}\tilde\Omega_m\,.
\ea
\subsection{Second moment}

The special functions related to the second moment of the Boltzmann equation are

\be
{\cal I'} (\hat{m}) \equiv \hat{m} \, \partial_{\hat{m}}{\cal I}=\hat{m}^2 \int d\hat{p} \,   \frac{\hat{p}^4}{\sqrt{\hat{p}^2 + \hat{m}^2}} f' \!\left(\!\sqrt{\hat{p}^2 + \hat{m}^2}\right) ,
 \label{eq:matching2}
\ee
and
\be
{\cal I}'_q (\hat{m}) \equiv \partial_{q}{\cal I}= \int d\hat{p} \,  \hat{p}^4  f'_q \!\left(\!\sqrt{\hat{p}^2 + \hat{m}^2}\right).
 \label{eq:matching3}
\ee
%

\bibliography{Tsallis}

\end{document}